\documentclass[%
reprint,amsmath,amssymb, longbibliography,aps,prd, nofootinbib,]{revtex4-1}

\usepackage{graphicx}% Include figure files
\usepackage{bm}% bold math
\usepackage{orcidlink}
\usepackage{mathrsfs}  
\usepackage{xcolor}  
\usepackage{upgreek}
\usepackage[normalem]{ulem}
\usepackage{cancel}

\def\Rvsd#1{\textcolor{red}{#1}}
\def\Rvsd#1{\textcolor{black}{#1}}

\def\dd#1#2{_{#1#2}}
\def\uu#1#2{^{#1#2}}
\def\be{\begin{equation}}
\def\ee{\end{equation}}
\def\bea{\begin{eqnarray}}
\def\eea{\end{eqnarray}}
\def\bpm{\begin{pmatrix}}
\def\epm{\end{pmatrix}}
\def\nnb{\nonumber}

\def\half{{\textstyle \frac 12}}

\def\quarter{{\textstyle \frac 14}}

\def\vac#1{{\bf#1}}
\def\bnabla{{\pmb{\nabla}}}

\def\df#1{{\sf#1}}

\DeclareMathOperator{\Grad}{{\mathop\mathbf{Grad}\ \! }}
\DeclareMathOperator{\Div}{{\mathop\mathrm{Div}\ \! }}
\DeclareMathOperator{\Curl}{{\mathop\mathbf{Curl}\ \! }}

\begin{document}

\title{Classical Kalb-Ramond field theory in curved spacetimes}

\author{Bertrand Berche \orcidlink{0000-0002-4254-807X}} 
\email{bertrand.berche@univ-lorraine.fr}
\affiliation{Laboratoire de Physique et Chimie Th\'eoriques,\\
 Universit\'e de Lorraine, CNRS, Nancy, France}
 \affiliation{${\mathbb L}^4$ Collaboration \& Doctoral College for the 
Statistical Physics of Complex Systems,\\
Leipzig-Lorraine-Lviv-Coventry}
\author{S\'ebastien Fumeron}
\affiliation{Laboratoire de Physique et Chimie Th\'eoriques,\\
 Universit\'e de Lorraine, CNRS, Nancy, France.}
\author{Fernando Moraes \orcidlink{0000-0001-7045-054X}}
\affiliation{Departamento de F\'{\i}sica,\\ Universidade Federal Rural de Pernambuco, 
 Recife, PE, Brazil}
\date{March 2021}

\begin{abstract}
We propose to develop the Kalb-Ramond theory in four-dimensional spacetime at the level of a classical field theory by following the same formal development steps as in Maxwell theory of standard electrodynamics.
Solutions of Kalb-Ramond theory in the presence of static sources in various curved spacetimes are then analyzed.  A question that we address here is that of a possible Kalb-Ramond polarization in curved spacetimes, like one can encounter a dielectric polarization in ordinary Maxwell electrodynamics in certain types of spacetimes.
\end{abstract}

\maketitle

\section{Introduction}
Electrodynamics is a remarkable theory. Synthesized in a set of four compact equations by Maxwell in the late nineteenth century, it is in a way a model of physical theory on which generations of physicists have subsequently developed new theories, to deal with gravity or fundamental interactions in the Standard Model.

Its applications are innumerable~\cite{zangwill_2012}, but is Maxwell theory, this model of physical theory  definitely set in stone?
Physicists never stop trying to circumvent, generalize, unify, existing theories and the same is true for electrodynamics. Maxwell theory has many mathematical and physical properties. It is linear,  gauge covariant, invariant under time-reversal symmetry, Abelian,  Lorentz invariant, can even be made generally covariant, it has Bianchi identities built in, etc, and one may wish to preserve or not these properties or some of them, when one tries to elaborate a new theory. Many examples already exist. Born-Infeld theory relaxes the linearity constraint, and even the time reversal or parity symmetries in some of its extensions,  Proca theory is not gauge invariant, 
dual electrodynamics does not obey Bianchi identities, Yang-Mills theories, viewed as generalization of Maxwell theory, are not Abelian, etc.

Kalb-Ramond (KR) theory~\cite{Kalb1974classical} is a beautiful alternative to Maxwell theory. Not in the sense that it has something to say on the fields which intervene in electrodynamics, but as it keeps most (if not all) the previous properties mentioned to build a new theory which is likely to capture some other reality of our world.
It was originally introduced as an interaction between extended objects in the context of superstring theory, 
with an action of the form
\bea
S_{\rm string}&=&-\mu^2\int (d\sigma\dd\mu\nu d\sigma\uu\mu\nu)^{1/2}-g_0\int d\sigma\dd\mu\nu B\uu\mu\nu\nnb\\ &&\qquad-\int d^4x
{\textstyle\frac 1{12}}M_{\lambda\mu\nu}M^{\lambda\mu\nu}
\eea
where $d\sigma\uu\mu\nu$ is the area element of the superstring's world sheet and $M_{\lambda\mu\nu}$ a third-rank antisymmetric tensor which derives from a gauge field $B\dd\mu\nu$. 

Kalb-Ramond theory also attracted the attention of the gravitation community as it appeared as a candidate for the torsion field \cite{doi:10.1080/00107519508222143,Hammond_1999,Hammond_2002}.
In a general study of connected spaces (spacetimes in the context of Physics), Elie Cartan introduced the concept of torsion which, together with the curvature, is a characteristic of the connection. He tried very early to attract the attention of Einstein (see \cite{10.2307/j.ctt13x1bd2}) and this probably opened the era of unified field theories on which Einstein himself~\cite{balibar_1993}, and many others~\cite{Eddington_1963,schrodinger_1985}, have concentrated considerable efforts (authoritative monographs are e.g. \Rvsd{Refs.}~ \cite{lichnerowicz_1955,tonnelat_1965,Vizgin_1994,krasnov_2020}). The relevance of Kalb-Ramond theory is now pervading condensed matter physics, where emerging Kalb-Ramond fields have been found in quantum liquid crystals \cite{Beekman2017}, semiconductor-metal hybrids \cite{Rodriguez2020}, and fluids \cite{Matsuo2021}.

In this ``classical gravity'' context, we believe that a study of the classical Kalb-Ramond theory in curved spacetimes, say at the level of Maxwell equations, is missing in the literature and it is our aim to try to fill this gap. 
We will thus propose to follow the footsteps of Maxwell theory,
with only a minor extension as an initial prerequisite,
the \Rvsd{Kalb-Ramond} hypothesis that a second-rank antisymmetric tensor can play the role of a gauge field.
We will then develop the known machinery, jumping from the tensor formulation \Rvsd{to} differential forms formalism which has the great advantage of allowing a very compact formulation of Maxwell theory and of being described by coordinate-free equations, and ending with ordinary vector formalism to  highlight in a concrete and obvious way the differences with the Maxwell equations in their  original form.

We will also discuss \Rvsd{briefly} the possible role of Kalb-Ramond fields \Rvsd{as a candidate as torsion field} in the context of spacetime geometry,  and propose simple solutions for various symmetric spacetimes.

\section{A masterpiece of theory}

\subsection{Maxwell theory}

For that purpose, our starting model is Maxwell-Lorentz electrodynamics, whose structure we summarize to then transpose it to the KR theory. Interactions between electric sources are mediated by the electromagnetic field, which, in terms of exterior algebra, can be encapsulated within four postulates. The first two posit the existence of two closed forms in four-dimensional spacetime, a 2-form $\df F$ for the field and a 3-form ${\mathscr J}$ for the sources, that is 
\bea d\df F&=&0,\label{eq-1}\\
d {\mathscr J}&=&0,\label{eq-2}\eea where $d$ is the exterior derivative (for an account of exterior calculus in connection with electrodynamics, see \cite{fumeron2020improving}). These two forms are thus locally exact, meaning that there exists a 1-form $\df A$ and a 2-form $\df G$ such that 
\bea\df F&=&d\df A,\label{eq-3}\\
{\mathscr J}&=&d\df G.\label{eq-4}\eea
Equations (\ref{eq-1}) and (\ref{eq-4}) are Maxwell equations, the first one being the Bianchi identity \Rvsd{(or structure equation)}, and the second the ``equation of motion''. (\ref{eq-3}) is the definition of the field strength and (\ref{eq-2}) is the charge conservation equation. Translated into ordinary vector language, this leads (in flat spacetime) to four equations 
$\bnabla\times\vac E+\partial_t\vac B=0$, $\bnabla\cdot\vac B=0$, $\bnabla\cdot\vac D=\rho$ and $\bnabla\times\vac H-\partial_t\vac D=\vac j$,
involving four independent vector fields, $\vac E$, $\vac B$, $\vac D$ and $\vac H$ which enter the definitions of $\df F$ and $\df G$,
and sources $\rho$ and $\vac j$ which define the current 3-form ${\mathscr J}$. 

The system of equations is not closed, and one needs further hypotheses, the constitutive equations,  fixed by postulating an action,
\be
S_{\rm EM}[A]=-\int \half \mu_0^{-1}\df F\wedge\star\df F+\df A\wedge\star\df J.\label{eq-action-diffform}
\ee 
This expression is justified by the requirement that one needs, under the integral, a Lorentz invariant scalar quantity (for a Lorentz invariant theory), or more generally a scalar under arbitrary diffeomorphisms. Indeed,   the exterior product of a $p$-form $\df u$ with its dual (or with the dual of another $p$-form) $\df u\wedge\star\df u$ is such a quantity, proportional to a full index contraction $u_\mu u^\mu$ and to the volume-form $d\hbox{vol}$. Therefore, a term in $\df F\wedge\star\df F$ in  the field Lagrangian density, being quadratic in the gauge field derivatives, has the good properties. 
Also, rather than the 3-form $\mathscr J$, let us  mention that it is  customary to introduce in the action a current 1-form $\df J$: for the same reason as above, one is \Rvsd{then} confident in the fact that $\df A\wedge\star\df J$ is a scalar invariant multiplying the volume-form.

Furthermore, the free field term displays gauge invariance as it is obviously unchanged in the transformation \be\df A\to\df A+d\upchi,\label{eq-gaugeA}\ee 
with $\upchi$ a zero-form.
Owing to the charge conservation, this transformation also leaves the interaction term unchanged, up to boundary terms. Indeed, one can write the modification of this term under the gauge transformation (\ref{eq-gaugeA}) as $d\upchi\wedge\star\df J=d(\upchi\wedge\star\df J)+\upchi\wedge\star d^\dagger\df J$. The first term contributes to the action functional by a surface term which we can neglect by assuming that there are not sources at infinity and, using the properties of the coderivative $d^\dagger$, the second term is modified into $\upchi\wedge d{\mathscr J}=0$, where use was made of equation (\ref{eq-2}) and the relation ${\mathscr J}=-\star\df J$, still to be shown.
Note that the sign of the coupling between the charges and the gauge field is essentially arbitrary  and we follow here reference~\cite{bertlmann_1996}.  

Being written in differential forms formalism, the theory is linear, coordinate-free, and displays general covariance. All these properties make it a ``pleasant'' theory.

Minimization of the action (\ref{eq-action-diffform}) leads to the equation of motion in one of the  following forms, in terms of the exterior derivative, $d$, or of the coderivative, $d^\dagger$,
\be  \mu_0^{-1}d^\dagger\df F=-\df J\quad\hbox{or}\quad \mu_0^{-1} d\star\df F=-\star\df J.\label{EL-Maxwell-form}\ee
The relation among  \Rvsd{the current 1-form $\df J$ and the current 3-form ${\mathscr J}$} is just ${\mathscr J}=-\star\df J$, as one can read from comparing (\ref{eq-4}) and (\ref{EL-Maxwell-form}), and this also provides the identification of
\be
\df G=\mu_0^{-1}\star\df F \label{M-const}
\ee
which is precisely the constitutive relation announced and which closes the system of equations.

A complete theory for electrodynamics \Rvsd{also} requires \Rvsd{the dynamics of the sources given by the Lorentz force, but we don't need it for the present purpose.} 
 It must be remarked that despite Maxwell theory being 150-year-old, the question of its foundations is still debated and our choice of axioms is not unique (for other proposals, see for instance \cite{hughes1992feynman,ObukhovHehl,sobouti2015lorentz,heras2016axiomatic}).

One might appreciate having the link with the tensor form of Maxwell equations. 
The action in this formalism,
\bea
S_{\rm EM}[A]&=&\int  d^4x\ \! {\mathscr L}_{\rm EM}\nnb\\
&=&-\int  d^4x\ \!\sqrt{-g}\ \!\Bigl(\quarter \mu_0^{-1} F\dd\mu\nu F\uu\mu\nu+A_\mu j^\mu\Bigr)\label{eq-action-tenseur}
\eea
is probably more usual. It obeys the requirements of being covariant under diffeomorphisms, gauge covariant, and invariant under parity reversal (this latter condition avoids a contribution in  $F\dd\mu\nu {\mathscr F}\uu\mu\nu$ with the dual field strength tensor ${\mathscr F}\uu\mu\nu$ introduced below). It follows that
\be
-\frac{\partial{\mathscr L}_{\rm EM}}{\partial A_\rho}=\sqrt{-g} j^\rho
\ee
which suggests, together with the Euler-Lagrange equation, to define the constitutive relation in the form of
\be
G\uu\sigma\rho=-\frac{\partial{\mathscr L}_{\rm EM}}{\partial(\partial_\sigma A_\rho)}
=\mu_0^{-1}\sqrt{-g}F\uu\sigma\rho\label{eq-constitutive-tensor}
\ee
for the equation of motion to be
\be
\frac{1}{\sqrt{-g}}\partial_\sigma G\uu\sigma\rho=\frac 1{\mu_0}\frac{1}{\sqrt{-g}}\partial_\sigma
(\sqrt{-g}F\uu\sigma\rho)=j^\rho\label{eq-motion-tensor}\ee
with the conservation of the electric charge 
\be
\partial_\rho(\sqrt{-g}j^\rho)=0
\ee
imposed by the antisymmetry of $G\uu\sigma\rho$.

In tensor form, it is common to introduce the dual field strength tensor
\be
{\mathscr F}\uu\mu\nu=\half\epsilon^{\mu\nu\rho\sigma}F\dd\rho\sigma
\ee
(with $\epsilon^{\mu\nu\rho\sigma}$ the Levi-Civita tensor)
in terms of which the Bianchi identity reads as (in Minkowski spacetime $\mathbb M^4$)
\be
\partial_\mu{\mathscr F}\uu\mu\nu=0.\label{eq-Bianchi-tensor}
\ee
From the similarity of equations  (\ref{eq-1}) and (\ref{eq-Bianchi-tensor})
on one hand, and of (\ref{eq-4}) and (\ref{eq-motion-tensor}) on the other hand, it is tempting to associate $\df F$ to ${\mathscr F}\uu\mu\nu$ and $\df G$ to $G\uu\mu\nu$. However, this is misleading, because  (\ref{eq-1}) and (\ref{eq-4})
are in fact three-indices equations, e.g. $d\df F=\frac1{3!}(d\df F)_{\lambda\mu\nu}dx^\lambda\wedge dx^\mu\wedge dx^\nu$. Hence, the proper form of the Bianchi equation is better written as a third rank tensor equation
\be
\partial_\lambda F\dd\mu\nu+\partial_\mu F\dd\nu\lambda+\partial_\nu F\dd\lambda\mu=0
\ee
which is true, even in curved spacetimes, as soon as one has defined the ``electromagnetic curvature'' as
\be F\dd\mu\nu=\partial_\mu A_\nu-\partial_\nu A_\mu.\label{eq-FvsA}\ee
This being said, it is now  clear that $\df F$ corresponds to $F\dd\mu\nu$.
Accordingly, we can \Rvsd{also} rewrite the equation of motion \Rvsd{as a third rank tensor form}, introducing the dual of $G\dd\mu\nu$,
\be
{\mathscr G}\uu\mu\nu=\half\epsilon^{\mu\nu\rho\sigma}G\dd\rho\sigma,
\ee
as
\be
\partial_\lambda {\mathscr G}\dd\mu\nu+\partial_\mu {\mathscr G}\dd\nu\lambda+\partial_\nu {\mathscr G}\dd\lambda\mu=-\sqrt{-g}{\mathscr J}_{\lambda\mu\nu}.
\ee
This equation is not as usual as (\ref{eq-motion-tensor}), but
this again shows  the correct correspondence, which is between $\df G$
and ${\mathscr G}\dd\mu\nu$.

Up to now, we have implicitly assumed a certain number of conventions. It is probably time  to become more explicit with the choices used in this paper.
We consider objects (tensors) defined on a 4-dimensional metric manifold ${\cal M}$ 
equipped with a metric tensor $g\dd\mu\nu$. We use the signature convention
with $g\dd 00\ge 0$,  the sign of the antisymmetric Levi-Civita tensor is chosen such as $\epsilon_{0123}=+1$ in $\mathbb M^4$ (where it coincides with the Levi-Civita symbol) and the orientation of the volume-form  is $d\hbox{vol}=\sqrt{-g}\ \!dx^0\wedge dx^1\wedge dx^2\wedge dx^3$ with $g=\hbox{det}\ \!(g\dd\mu\nu)$. Differential forms are written in sanserif Latin or upright Greek, and their Hodge dual in calligraphic.
 In this paper, we also mostly follow the notations and conventions of the excellent books {\em Differential geometry, gauge theories, and gravity}  by G\"ockeler and Sch\"ucker~\cite{gockeler_schucker_1987} and Bertlmann's {\em Anomalies in quantum field theory}~\cite{bertlmann_1996}. In particular, it means that we use the Hodge star operation such that it discriminates the forms according to their degree as $\star\star\df u=-\df u$ for even degree while
$\star\star\df u=\df u$ for odd degree, and the coderivative is  simply $d^\dagger= \star\ \! d\ \! \star $. Among the other implicit choices, we should specify that the vector fields $\vac E$, \Rvsd{$\vac B$, $\vac D$ and $\vac H$} have their usual meaning if we define
% $\overset 2F=E_i cdt\wedge dx^i-\half B_i\epsilon_{ijk}dx^j\wedge dx^k$ and $\overset 2G=-\half cD_i\epsilon_{ijk}dx^j\wedge dx^k-H_i cdt\wedge dx^i$ with $i,j,k$ ordinary Euclidean indices. Similarly, $\overset 3{\mathscr J}=-\rho dx^1\wedge dx^2\wedge dx^3+\half j_i\epsilon_{ijk}cdt\wedge dx^j\wedge dx^k$.
$\df F=-\df E\wedge dt-\df B$ and $\df G=-c\df D+\df H\wedge cdt$ with $\df E$,  \Rvsd{$\df B$, $\df D$ and $\df H$} differential forms in
3-dimensional space. The signs are imposed by the signature choice. Similarly, the current 3-form is ${\mathscr J}=-\rho c+\df j\wedge cdt$. We have decided, contrary to  common use, to keep the constants $c$ \Rvsd{and} $\mu_0$, to make explicit the difference between $F\uu\mu\nu$ and $G\uu\mu\nu$, even in flat spacetime.

\subsection{Kalb-Ramond theory}

A very instructive generalization of Maxwell electrodynamics is provided by Kalb-Ramond extension to higher-order tensors \Rvsd{(a succinct treatment of the generalization to higher order tensors and arbitrary spacetime dimensions can be found in the book of Ort\'\i n, \cite{Ortin})}. This is an example which illustrates the introduction (besides, or instead of, the Faraday tensor $F\dd\mu\nu$), of new fields, $M_{\lambda\mu\nu}$, which couple to new types of matter sources.

Let us thus assume the existence of a field strength 3-form $\df M$ and a generalized conserved charge 2-form acting as a source term ${\mathscr S}$, both forms being closed,
\bea d\df M&=&0,\label{eq-17}\\
d {\mathscr S} &=&0.\label{eq-18}\eea 
Therefore, there exists a gauge field 2-form $\df B$ and a 1-form $\df N$ such that 
\bea\df M&=&d\df B,\label{eq-19}\\
{\mathscr S}&=&d\df N.\label{eq-20}\eea
Equations (\ref{eq-17}) and (\ref{eq-20}) are equivalent to Maxwell equations (\ref{eq-1}) and (\ref{eq-4}).
We are now facing the same problem as with Maxwell theory, and we need some constitutive relations to close the system of equations.
For that purpose, let us assume an action functional given by an obvious generalization of (\ref{eq-action-diffform}),
\be
S_{\rm KR}[B]=-\int \half {g_0}^{-1}\df M\wedge\star\df M+\df B\wedge\star\df S\label{eq-action-diffformKR}
\ee 
with ${g_0}$ an unknown dimensional parameter and $\df S$ introduced in the action to describe the sources in analogy with the Maxwell case.
The variation of $S_{\rm KR}[B]$ leads to 
\bea
\delta S_{\rm KR}[B]&=&-
\int\delta\df B\wedge({g_0}^{-1}\star d^\dagger\df M+\star\df S)+\hbox{BT}
\eea
such that, up to the boundary terms (BT), the equation of motion follows in either of the two forms
\be
d^\dagger\df M=-{g_0}\df S\quad\hbox{or}\quad d\star\df M=-{g_0}\star\df S.\label{EL-KR-form}
\ee
The analogy with (\ref{EL-Maxwell-form}) is transparent.
Using the duality relation for the sources, ${\mathscr S}=-\star\df S$, the constitutive relation follows,
\be
\df N={g_0}^{-1}\star\df M.
\ee

Although the approach using differential forms has the advantage of a high degree of generality,  \Rvsd{valid} in particular \Rvsd{in} the case of curved spacetimes, it is convenient to write also these expressions in their tensorial forms, or even in a very traditional vector shape (this was briefly done in \Rvsd{flat spacetime in} the Appendix of the original paper~\cite{Kalb1974classical}).
Thus, we have an antisymmetric third rank field strength tensor $M_{\lambda\mu\nu}$ which derives from an antisymmetric second-rank gauge tensor $B\dd\mu\nu$
\be M_{\lambda\mu\nu}
=\partial_\lambda B\dd\mu\nu+\partial_\mu B\dd\nu\lambda +\partial_\nu B\dd\lambda\mu.\label{eq_KRH}\ee
$M_{\lambda\mu\nu}$ has four independent components while  $B\dd\mu\nu$ has six.
This definition leads to the Bianchi equation
 \be
 \partial_\kappa M_{\lambda\mu\nu}-\partial_\lambda M_{\mu\nu\kappa}+\partial_\mu M_{\nu\kappa\lambda}
 -\partial_\nu M_{\kappa\lambda\mu}=0.\label{eq-BianchiKRtensor}
 \ee
Gauge invariance of the Kalb-Ramond theory is guaranteed by the following transformation of the gauge field
\be
B\uu\mu\nu\to {B'}\uu\mu\nu=B\uu\nu\mu+\partial^\mu\xi^\nu-\partial^\nu\xi^\mu
\ee
as it can easily be shown.

The free field Lagrangian density  is  built in analogy to the Maxwell case, $-\frac 1{12} {g_0}^{-1}M_{\lambda\mu\nu}M^{\lambda\mu\nu}$. The coupling of the gauge field $B\dd\mu\nu$ to external sources requires an antisymmetric second rank tensor describing the matter currents  to saturate the Lorentz indices, $-\half B\dd\mu\nu S\uu\mu\nu$. Such sources describe extended objects and Kalb-Ramond theory has a natural application in the context of string theories as we said in the introduction. 
The Kalb-Ramond action \index{Kalb-Ramond action} reads as
\bea
S_{\rm KR}[B]&=&\int d^4x\ \!{\mathscr  L}_{\rm KR}\nnb\\
&=&-\int d^4x\sqrt{-g}\ \!\left(
{\textstyle\frac 1{12}}{g_0}^{-1} M_{\lambda\mu\nu}M^{\lambda\mu\nu}+ \half B\dd\mu\nu S\uu\mu\nu
\right).\nnb\\ \label{eq_KRAction}
\eea

The Euler-Lagrange equation is easily applied to (\ref{eq_KRAction}), leading to the equation of motion
\be
\frac 1{g_0}\frac{1}{\sqrt{-g}}
\partial_\sigma (\sqrt{-g}M^{\sigma\mu\nu})=S\uu\mu\nu,
\label{eq_KREL}
\ee
where the antisymmetry (\ref{eq_KRH}) of the Kalb-Ramond  field strength  tensor has been used.

This equation of motion, together with the  antisymmetry of $B\uu\mu\nu$,
leads to the conservation of the second-rank tensor current,
\be\partial_\mu (\sqrt{-g}\ \! S\uu\mu\nu)=0.\ee

Rephrased now in the language of ordinary functions and vectors, we can write equations that look very similar to the usual Maxwell equations. \Rvsd{Let us introduce an effective $3$-metric $\gamma\dd ij$ (with $i,j=1,2,3$) via a space-time decomposition
\be
ds^2 =g\dd\mu\nu dx^\mu dx^\nu=g\dd 00(dx^0+(g\dd 0i/g\dd00)dx^i)^2-\gamma\dd ijdx^idx^j\ee
with $\gamma\dd ij=-g\dd ij+g\dd 0i g\dd 0j/g\dd 00$ and $-g=\gamma g\dd00$ with $g$ the $4$-spacetime metric determinant and $\gamma$ that of the $3$-metrics.
}
We define a density $\kappa$ (for ``Kalb'') and a pseudo-vector $\vac R$ (for ``Ramond'') corresponding to two purely spatial differential forms, respectively a 3-form $\df K$ and a 2-form $\df R$ such that $\df M=\df R\wedge cdt+\df K$ (here we do not \Rvsd{impose} negative signs like in the Maxwell case since there are no standardized definitions of the fields $\vac R$ and $\kappa$).
This enables us to write 
the second Bianchi equation as two equations, $d_3\df R+\frac 1c\partial_t\df K=0$ and $d_3\df K=0$ (here $d_3$ is the purely spatial exterior derivative).
\Rvsd{The components of $\kappa$ and $\vac R$ are obtained from the identification $\kappa=\frac 16\frac{\epsilon^{ijk}}{\sqrt\gamma}K_{ijk}$ with $K_{ijk}=M_{ijk}$ 
and $R^i=\frac 12\frac{\epsilon^{ijk}}{\sqrt\gamma}M_{0jk}$. 
The first Bianchi identity is then written as 
\be \frac 1c\frac 1{6\sqrt\gamma}\epsilon^{ijk}\partial_t K_{ijk}-\frac 1{\sqrt\gamma}\partial_i(\sqrt\gamma R^i)=0.\ee
In vector form, we obtain an expression which looks very similar to that of flat spacetime \cite{Kalb1974classical},
\be
\frac 1c{\mathscr D}_t{\kappa}-\Div \vac R=0,\label{eq-KR_Rkappa}
\ee
with here the divergence operator defined in curved space \cite{LandauLifshitz} as $\Div \vac R=(1/\sqrt\gamma)\partial_i(\sqrt\gamma R^i)$ and the time derivative as ${\mathscr D}_t{\kappa}=(1/\sqrt\gamma)\partial_t (\sqrt\gamma\kappa)$.}
\Rvsd{The second equation following from Bianchi identity written in $3$-dimensional forms}  does not have any counterpart there, except that $\kappa$ is a density.

The inhomogeneous equations also have their vector counterparts, 
  calling for the introduction of two additional  fields, a vector field $\vac T$ and a scalar field ${\lambda}$, associated respectively to a 1-form $\df T$ and a zero-form
  $\uplambda$ which constitute $\df N=g_0^{-1}\star\df M=\df T+\uplambda\wedge cdt$.
  The  external 
 sources ${\mathscr S}=\upsigma+\df s\wedge cdt$ are represented by a pseudo-vector $\pmb\sigma$ (corresponding to the 2-form $\upsigma$) and a vector, $\vac s$ (the 1-form $\df s$). We get then the equations of motion
 $d_3\df T=\upsigma$
 and $d_3\uplambda+\frac 1c\partial_t\df T= \df s$.
  In terms of ordinary fields $\vac T$ and  ${\lambda}$, one has
\bea
&&
\Rvsd{
\Curl \vac T=\pmb{\sigma}
},\label{eq-rotS}\\
&&
\Rvsd{
\frac 1c\mathscr D_t\vac T-\Grad{\lambda}={\vac s}
},\label{eq-gradL}
\eea
themselves analogs of
\bea
&&\bnabla\cdot\vac D=\rho,\label{eq-MaxwDvector}\\
&&\bnabla\times\vac H-\partial_t\vac D=\vac j\label{eq-MaxwHvector}
\eea
in Maxwell theory. \Rvsd{The components of the vector $\vac T$ are  those of the one-form $\df T$, 
$T_i=\frac1{2g_0}\sqrt\gamma\sqrt{g\dd00}\epsilon_{ijk}M^{0jk}$ and the scalar field $\lambda$ is defined as $\lambda=-\frac1{6g_0}\sqrt\gamma\sqrt{g\dd00}\epsilon_{ijk}M^{ijk}$. 
The components of the sources  $\pmb\sigma$ are  $\pmb\sigma|^i=\sqrt{g\dd 00}s\uu 0i$  and those of $\vac s$ are built from the space components of the two-form $\df s$, 
$\vac s|_i=\sqrt{g\dd 00}s_i=\frac 12\sqrt{g\dd00}\sqrt\gamma\epsilon_{ijk}{\mathscr S}\uu jk$. The curl of a vector field in curved space is defined as \cite{LandauLifshitz}
\be\left.\Curl \vac T\right|^i=(1/\sqrt\gamma){\epsilon^{ijk}}\partial_j T_k,\ee  and the gradient operator is defined ordinarily as $\left.\Grad{\lambda}\right|_i=\partial_i\lambda$.}

\Rvsd{
There is an interesting observation made by Landau and Lifshitz \cite{LandauLifshitz} in the case of Maxwell electrodynamics:
\begin{quotation}
The reader should note the analogy (purely formal, of course) with Maxwell equations for the electromagnetic field in material media. In particular, in a static gravitational field the quantity $\sqrt\gamma$ drop out of the terms containing time derivatives (\dots) We may say that with respect of its effect on the electromagnetic field a static gravitational field plays the role of a medium with electric and magnetic permeabilities $\epsilon=\mu=1/\sqrt{g\dd 00}$.
\end{quotation}
It also simplifies the problem here, because then most of the flat spacetime formulas of vector analysis apply and the differential operators  ${\mathscr D}_t$ and $\Curl$ or $\Grad$ commute with each other.}

The conservation of the Kalb-Ramond charges follows from the equations of motion and reads as $d_3\upsigma=0$ and $d_3\df s+\frac 1c\partial_t\upsigma=0$.  In vector formalism, equation (\ref{eq-rotS}) demands that $\pmb\sigma$ is divergence free and, taking the curl of (\ref{eq-gradL}), we get a continuity equation of the form (\ref{eq61})
\bea 
&&\Rvsd{\Div }\pmb\sigma=0,\\
&&\Rvsd{\Curl }{\vac s}\Rvsd{-}\frac 1c\partial_t{\pmb\sigma}=0.\label{eq61}
\eea 

\Rvsd{The definition of gauge potentials is suggested by the} form of the Bianchi equations above. \Rvsd{They}  suggest the introduction of a potential 
2-form $\upalpha$ and a potential 1-form $\upphi$ such that $\df K=d_3\upalpha$ and $\df R=-d_3\upphi-\frac 1c\partial_t\upalpha$. 
In vector notation, \Rvsd{these} translate into two gauge vector potentials $\pmb\alpha$ and $\pmb\phi$  such that  the Kalb-Ramond fields are defined in terms of these as
\bea
&&
\Rvsd{\kappa=\Div \pmb\alpha,}
\\
&&
\Rvsd{
\vac R=-\Curl \pmb\phi-\frac 1c\partial_t\pmb\alpha,
}
\eea
\Rvsd{with $\alpha^i=\frac 12\frac{\epsilon^{ijk}}{\sqrt\gamma}\alpha\dd jk$ the components of the vector $\pmb\alpha$ in terms of those of the $2$-form $\upalpha$.}

We know that in vector notation, Maxwell theory has in the vacuum, and in the absence of a gravitational field, the nice property that there is a simple proportionality  between $\vac E$ and $\vac D$ on one hand and between $\vac B$ and $\vac H$ on the other hand, these are the  constitutive relations in 3$D$ formalism.
Therefore, the four coupled first-order partial differential  Maxwell equations decouple into two second-order equations known as wave equations. 
This property in fact hides a duality relation in terms of forms.
The same strategy can be used in Kalb-Ramond theory, wherefrom ${g_0}^{-1}\star (\df R\wedge cdt) =\df T$ and ${g_0}^{-1}\star\df K=\uplambda c dt$, and one deduces the constitutive relations in the form
\bea
{g_0}^{-1}\vac R=\vac T,\label{eq-ConstRS}\\
{g_0}^{-1}\kappa=\lambda.\label{eq-Constkappasigma}
\eea 
This leads to the Kalb-Ramond equations, now written only in terms of the fields $\vac R$ and $\kappa$,
\bea
&&\Rvsd{\Div}\vac R\Rvsd{-}\frac 1c\partial_t{\kappa}=0,\label{eq-KRRK}\\
&&\Rvsd{\Curl}\vac R={g_0}\pmb\sigma,\label{eq-KRRsigma}\\
&&\Rvsd{\Grad\kappa}\Rvsd{-}\frac 1c\partial_t\vac R={g_0}\vac s.\label{eq-KRKRl}
\eea 
One may wonder why only three equations are required in Kalb-Ramond theory whereas four are needed in Maxwell's (once constitutive relations have been used). This is a natural consequence of Helmholtz decomposition theorem: Maxwell theory involves two dynamic vector fields (each known from its curl and divergence), whereas Kalb-Ramond's involves one dynamic vector field and one dynamic scalar field.

Taking the time partial derivative of (\ref{eq-KRRK}) and using (\ref{eq-KRKRl}) we get, 
\Rvsd{for the scalar field $\kappa$,}
 a non-homogeneous wave equation 
 \be 
\Rvsd{\Delta_{\rm LB}}{\kappa}-\frac{1}{c^2}\partial_t^2{\kappa}=\Rvsd{g_0}\bnabla\cdot\vac s,\label{eq-prop1}
\ee 
\Rvsd{with the Laplace-Beltrami operator
\be
\Delta_{\rm LB}\kappa=\Div(\Grad\kappa)=\frac{1}{\sqrt\gamma}\partial_i(\sqrt \gamma \gamma\uu ij\partial_j\kappa).
\ee}
\Rvsd{The wave equation takes the standard form} $\Box\kappa=0$ in \Rvsd{flat} empty space, meaning that the scalar field propagates in the form of waves at the light velocity $c$. 
Now, taking the gradient of (\ref{eq-KRRK}),  the curl of (\ref{eq-KRRsigma}), and combining with (\ref{eq-KRKRl}) also leads to a non-homogeneous wave equation for the vector field $\vac R$,
\bea 
\Rvsd{\Grad(\Div\vac R)-\Curl(\Curl\vac R)}&-&\frac{1}{c^2}\partial_t^2\vac R\nonumber\\
&&=\Rvsd{-}g_0(\bnabla\times{\pmb\sigma}+\frac 1c\partial_t\vac s).\nonumber\\\label{eq-prop2}
\eea 
Again, one retrieves $\Box\vac R=0$ in source-free \Rvsd{flat} space.

\section{The link with totally antisymmetric torsion}
The route for geometric studies of spacetime structure was opened by Albert Einstein with his General Theory of Relativity. It was there that he developed the principle of general covariance. This principle would be called upon to become one of the foundations of Physics, not only in the fields of gravitation and electromagnetism but also more generally with the advent of gauge theories which will revisit this question.
General covariance (under arbitrary changes of coordinates) is naturally expressed in tensorial form and the laws of dynamics, which involve fields and their derivatives, have to face the problem that, in general, a field derivative {\em is not} a tensor. This is repaired by the introduction of a connection ${\Gamma^\sigma}\dd\mu\nu$, which adds a piece to the derivative to make it a tensor. Consider for example a vector field $V^\sigma$. While $\partial_\mu V^\sigma$ is not a (mixed) second-rank tensor, the quantity $\partial_\mu V^\sigma+{\Gamma^\sigma}\dd\mu\nu V^\mu$ is such a tensor, provided that ${\Gamma^\sigma}\dd\mu\nu$ satisfies appropriate transformation laws under a change of the coordinates.
In the study of gravitation, Einstein considered metric spacetimes and has chosen to specify the Levi-Civita connection, given by the Christoffel symbols (in terms of the derivatives of the metric tensor $g\dd\mu\nu$).
This connection has the characteristic feature to be symmetric w.r.t the exchange of its two lower indices, 
 ${\Gamma^\sigma}\dd\mu\nu= {\Gamma^\sigma}\dd\nu\mu$, but this is not a necessary condition for a connection.
 When this is not the case, the antisymmetric part of the connection defines the torsion field
 \be
{S^\sigma}\dd\mu\nu= {\Gamma^\sigma}\dd\nu\mu- {\Gamma^\sigma}\dd\mu\nu.
 \ee
 
The torsion is, by construction, an antisymmetric tensor.
It comprises 24 independent components and can be separated into irreducible pieces~\cite{PhysRevD.19.3524}, a  vector $S_\mu={S^\sigma}\dd\sigma\mu$, a  pseudo-vector
$A_\rho=\varepsilon_{\rho\sigma\mu\nu}{S^{\sigma}}\dd\mu\nu$ and the remaining 16 components
are stored in a third-rank tensor $b_{\mu\nu\sigma}$. Various geometric theories incorporate torsion~\cite{doi:10.1080/00107519508222143}, and the one which is relevant for us is the torsion which derives from a second-rank antisymmetric gauge field, i.e., \Rvsd{following Hammond (e.g. in \cite{doi:10.1080/00107519508222143}), we assume a spacetime for which the torsion field obeys Kalb-Ramond equations,}
\be
S_{\lambda\mu\nu}=M_{\lambda\mu\nu}\label{eq-torsionKR}
\ee
as given in equation~(\ref{eq_KRH}) \Rvsd{possibly up to proportionality factors}.  \Rvsd{In this approach, the source of torsion is} the spin of elementary particles. A pretty nice feature of this theory is that contrary to \Rvsd{the more common} contact interaction which does not propagate~\cite{RevModPhys.48.393}, \Rvsd{if} torsion \Rvsd{is given by (\ref{eq-torsionKR})}, as we have seen in equations~(\ref{eq-prop1}) and (\ref{eq-prop2}),
\Rvsd{it}  obeys wave equations.

\section{Kalb-Ramond static field in curved spacetimes \label{Curvedsection} }
The Kalb-Ramond fields can be obtained via tensor calculus (\ref{eq-BianchiKRtensor}) and (\ref{eq_KREL}), via ordinary vector formalism (\ref{eq-KR_Rkappa}), (\ref{eq-rotS}) and (\ref{eq-gradL}), but in order to illustrate the power of exterior differential calculus, we will consider the static Kalb-Ramond field created by a localized source distribution in various spacetimes, assimilated to fixed backgrounds, \Rvsd{using mainly  Eqs.~(\ref{eq-17}) and (\ref{eq-20})}. \Rvsd{The source of the Kalb-Ramond field is supposed to be independent of the source of the gravitational field which is at the origin of the geometry of each spacetime considered.}
The question that we address here is the possibility of a non-trivial vacuum polarization, i.e. a non-trivial relation between the components of the dual form $\df T$ and those of $\df R$.
Before starting the calculation, a comment is needed on the definition of the form $\df R$.
Following the standard definition of the Faraday 2-form in electrodynamics, we have defined the Kalb-Ramond 3-form as $\df M=\df R\wedge cdt +\df K$. This is a natural definition in Minkowski spacetime, 
but in an arbitrary manifold, $cdt$ does not necessarily have an unambiguous meaning, and, under a change of coordinates, if $\df M$ remains unchanged, e.g. $\df M=\frac {1}{3!}M_{\lambda\mu\nu}dx^\lambda\wedge dx^\mu \wedge dx^\nu=\frac {1}{3!}M_{abc}\df e^a\wedge \df e^b \wedge \df e^c$, this is not the case for $\df R$ for which a particular choice of coordinate $ct$ was done.
One could also define $\df R$ and $\df K$ such that 
\be \df M =\df R'\wedge \df e^0 +\df K'\label{eq_DfnR}\ee
where $\df e^0$ is the unit vector \footnote{We use the term ``vector'' meaning an element of a generic vector space, not an ordinary vector that would be in the tangent spacetime.} in the time-like direction of the cotangent spacetime in an orthonormalized basis of 1-forms. This coincides with the standard definition in Minkowski spacetime, but one has to emphasize the fact that $\df R'$ and $\df K'$ defined in equation (\ref{eq_DfnR}) do not, in general, coincide with the $\df R$ and $\df K$ which  follow from  $\df M=\df R\wedge cdt +\df K$.  We will give both expressions in the forthcoming examples. The components in the orthonormalized basis are called ``physical components'' by Hartle~\cite{hartle_2003}, but there, the Physics obviously appears to be compatible with Special Relativity, \Rvsd{by construction}. In particular, there is no vacuum polarizability of the Minkowski vacuum, so that we will refer to non-trivial polarizability in the non-inertial local coordinate system.

\subsection{Cosmic string in Schwarzschild spacetime}

Let us first contemplate the case of a cosmic string crossing the origin of a Schwarzschild blackhole spacetime. This is an example of a spacetime characterized by a singular curvature.
 
 We use the Schwarzschild coordinates $ct,r,\theta,\varphi$ and the line element
 \bea
 ds^2&=&\Bigl(1-\frac {2M}r\Bigr)c^2dt^2\nnb\\&& -\Bigl(1-\frac {2M}r\Bigr)^{-1}dr^2
 -r^2d\theta^2-\alpha^2r^2\sin^2\theta d\varphi^2\quad\label{eq-ds2SchwCS}
 \eea
 where $\alpha$ is the cosmic string defect-angle parameter ($\alpha^2=1-4G\mu/c^2$ with $\mu$ the string stress tensor amplitude).
 Instead of the local basis $(cdt,dr,d\theta,d\varphi)$, the tetrad formalism  allows to write the line element in an orthonormalized 
basis $(\df e^{0},\df e^{1},\df e^{2},\df e^{3})$,
\be
ds^2=(\df e^{0})^2-(\df e^{1})^2-(\df e^{2})^2-(\df e^{3})^2\label{eq-ds2dual-tetrad}
\ee
with Minkowski metric. Here, the coframe basis vectors are obtained by inspection, $\df e^0=\Bigl(1-\frac {2M}r\Bigr)^{1/2}cdt$, $\df e^1=\Bigl(1-\frac {2M}r\Bigr)^{-1/2}dr$, $\df e^2=rd\theta$, and $\df e^3=\alpha r\sin\theta d\varphi$.
Using the equivalent of Gauss law, (\ref{eq-rotS}), which, due to the form degree is more like an Amp\`ere law for Kalb-Ramond fields, we can write
 \be
\int_{\Sigma} d\df T=\int_{\partial\Sigma}\df T= \int_{\Sigma} \upsigma\label{eqGKR}
 \ee
 \Rvsd{and call $\Phi_\sigma$ this quantity, by analogy with a flux.  Let us mention again that this quantity is produced by some charge density $\upsigma$ which is not specified more, and in particular, which is not the source of the metric (\ref{eq-ds2SchwCS}).
Equation  (\ref{eqGKR}) is a counting procedure of the Kalb-Ramond charges crossing the compact domain $\Sigma$: as underlined by \cite{ObukhovHehl}, (\ref{eqGKR}) defines a purely topological quantity and must not depend explicitly on the spacetime geometry (physical parameters $M$ and $\alpha$).   
}
 Assuming cylindrical symmetry in the Minkowskian coframe, we choose for the surface $\Sigma$ a disk of radius $\rho=r\sin\theta$ with contour $\partial\Sigma$. The total KR charge crossing the disk  is characterized by \Rvsd{the} flux denoted as $\Phi_\sigma$. The 1-form $\df T$ can be expressed either on the local basis,  or using the dual tetrad basis $(\df e^{0},\df e^{1},\df e^{2},\df e^{3})$ where cylindrical symmetry imposes an expression of the form $\df T=T_3\df e^{3}$. Then, using the relation to the local basis, $\df e^3=\alpha\sin\theta d\varphi$, one gets
 \be
 \df T=T_3\df e^{3}=T_3\alpha r\sin\theta d\varphi=T_\varphi d\varphi
 \ee
with  $T_3=T_3(\rho)=T_3(r\sin\theta)$. 
Therefore one deduces from (\ref{eqGKR})
 \be
 T_3(r\sin\theta)=\frac{\Phi_\sigma}{2\pi\alpha r\sin\theta}=\frac{\Phi_\sigma}{2\pi\alpha\rho},\quad\hbox{and}\quad
 T_\varphi=\frac{\Phi_\sigma}{2\pi}.\label{eq-59}
 \ee
 
 Now, in absence of the field $\df K$, using the relations 
 \be \df M=\df R'\wedge \df e^0=\df R\wedge cdt=g_0\star\df N={g_0}\star \df T\ee
 and the expression of the Hodge dual in the local co-tetrad frame (with metric tensor there denoted as $\eta\dd ab$),
 \be
 \star \df T=\sqrt{|\eta|}{\eta}\uu33 T_3\epsilon_{3012}\df e^0\wedge \df e^1\wedge \df e^2,
 \ee
 we can read that
 \bea
 g_0\star\df T&=&g_0 T_3 \df e^1\wedge \df e^2\wedge \df e^0\nnb\\
 &=&g_0\frac{T_\varphi}{\alpha \sin\theta}dr\wedge d\theta\wedge cdt.\label{eq-63}
\eea
 The first line leads to the ``Minkwoskian" identification
 $R'\dd 12 = g_0 T_3$, while in the local basis, 
we can interpret our result as a non-trivial vacuum polarization, since we can write
\be
R\dd r\theta=g(r\sin\theta)T_\varphi=\frac {g_0}{\alpha \sin\theta}\frac{\Phi_\sigma}{2\pi}\label{eq-64}
\ee
\Rvsd{with $g(r\sin\theta)=\frac {g_0}{\alpha \sin\theta}$ the analogue of a relative permittivity.}

One may  wonder about the robustness of this result with respect to a change of local coordinates. For that purpose, we will now consider
 a cosmic string in a  Schwarzschild spacetime with  Israel  coordinates~\cite{mueller2010catalogue}. The main interest is that the metric tensor
 is not diagonal in this case.
The Schwarzschild \Rvsd{plus string} metric in Israel coordinates $(y,x,\theta,\varphi)$ reads
\bea
ds^2&=&-r_S^2\Bigl[
4dx\Bigl(
\frac{y^2dx}{1+xy} +dy
\Bigr)\nnb\\
&&\qquad+(1+xy)^2(d\theta^2+\alpha^2\sin^2\theta d\varphi^2)
\Bigr]
\eea
where $x$ and $y$ are linked to Schwarzschild coordinates via $t=r_S(1+xy+\ln(y/x))$ and $r=r_S(1+xy)$.
We have also added the cosmic string with defect-angle $\alpha$.
 The (opposite of the) metric tensor takes the form
 \be
 (-g\dd\mu\nu)=\bpm 0 & 2r_S^2 &0&0\\2r_S^2 &\frac{4r_S^2y^2}{1+xy}&0&0\\0&0&r_S^2(1+xy)^2&0\\0&0&0&r_S^2(1+xy)^2\alpha^2\sin^2\theta\epm.
 \ee
 and its determinant is
 \be -g=4r_S^8(1+xy)^4\alpha^2\sin^2\theta.\ee
The cotetrad vectors are given by
 \bea
 &&\df e^0=r_S\frac{\sqrt{1+xy}}{y}dy,\label{cotet0}\\
 &&\df e^1=r_S\left(\frac{2y}{\sqrt{1+xy}}dx+\frac{\sqrt{1+xy}}{y}dy\right),\label{cotet1}\\
 &&\df e^2=r_S(1+xy)d\theta,\label{cotet2}\\
 &&\df e^3=r_S(1+xy)\alpha\sin\theta d\varphi,\label{cotet3}
 \eea
 with again Eq.~(\ref{eq-ds2dual-tetrad}) satisfied by this choice.
 Note that we specify (by choice) that the $0$ coordinate is along $y$.
 Again, in the orthonormalized basis, one has
 \be
 \int \df T=\Phi_\sigma=\int T_3(\rho) \ \!\df e^3
 \ee
 which requires that
 \be
 T_3(\rho)=\frac{\Phi_\sigma}{2\pi r_S(1+xy)\alpha\sin\theta}=\frac{\Phi_\sigma}{2\pi\alpha\rho}
 \ee
and, like in Schwarzschild coordinates, $T_\varphi=\Phi_\sigma/(2\pi)$ in the coordinate basis.
The Hodge dual $\star \df T$ is the same as the first line of equation~(\ref{eq-63}) in the orthonormalized basis, but it delivers, in the present coordinate basis, the following expression
\be
R\dd x\theta=\frac{2g_0r_S^2}{\alpha\sin\theta}\frac{\Phi_\sigma}{2\pi},
\ee
\Rvsd{hence a different ``permittivity''.}
The consequences of the choice of coordinate basis will be commented in the discussion.

\subsection{Wiggly cosmic string}

Cosmic strings may have a structure, and the case of the ``wiggly'' cosmic string is particularly interesting for our purpose, since the line element acquires a further space dependence. 
The presence of wiggles indeed generates a far gravitational field contribution and averaging the effect of these perturbations along the string increases the linear mass density ${\tilde\mu}$ and decreases the string tension $\tilde{T}$, (with an equation of state ${\tilde\mu}\:\tilde{T}=\mu^2$), leading the wiggly string to exert a gravitational field at large distances. 
We can thus address here the question of the form of the vacuum polarization in such a case.

In the weak gravitational field approximation, the linearized line element in the presence  of a wiggly string oriented along the $z$-axis is given by~\cite{Vilenkin1990,Vachaspati1991,Vilenkin1994,PhysRevD.96.084047}:
\bea
	ds^2&=&\left(1+8\varepsilon\ln\left({\rho}/{\rho_0}\right)\right)c^2dt^2 
	-d\rho^2\nonumber\\&&\qquad-\alpha^2 \rho^2 d\varphi^2 -\left(1-8\varepsilon\ln\left({\rho}/{\rho_0}\right)\right)dz^2, \label{Vilenkinmetric}
\eea
where the conical deficit-angle associated to the string is now $\alpha^2=1-4G({\tilde\mu}+\tilde{T})/c^2$ (with $4G({\tilde\mu}+\tilde{T})/c^2 \ll 1$). The parameter $\varepsilon$  defines the excess of mass-energy density, $2\varepsilon=G({\tilde\mu}-\tilde{T})/c^2$. The value of $G(\tilde{\mu}+\tilde{T})$ accounts for the discrepancy between flat and conical geometries, whereas $G({\tilde\mu}-\tilde{T})$ accounts for the discrepancy between straight and wiggly strings. The constant $\rho_0$ denotes the effective string radius \cite{patrick1994}. We also consider the limit where $\varepsilon\ln\left({\rho}/{\rho_0}\right) \ll 1$.

The calculation of $\df T$ follows the same lines of reasoning as in the previous section, and one just has a slight modification due to the use of a different ordering of the local (cylindrical) coordinates, $(ct,\rho,\varphi,z)$, which now demands that $\df T=T_2\df e^2$, with $\df e^2=\alpha\rho d\varphi$, thus
\be
\df T=\frac{\Phi_\sigma}{2\pi\alpha\rho}\df e^2=\frac{\Phi_\sigma}{2\pi} d\varphi.\label{eq-79}
\ee
The dual follows,
\be\star \df T=\frac{\Phi_\sigma}{2\pi\alpha\rho} dz\wedge d\rho\wedge cdt,\ee
hence $R\dd z\rho=g_0T_\varphi/(\alpha\rho)$.

\subsection{Chiral cosmic string  spacetime}
Probably more interesting is the case of a  non-diagonal metric associated with non-zero torsion, axially  localized.
As an example, we consider a chiral cosmic string for which the line element is given by
\be
ds^2=c^2dt^2-d\rho^2-\rho^2 d\varphi^2 - (\beta d\varphi+dz)^2,
\ee
where $\beta$ is the Burgers parameter of the cosmic string. \Rvsd{Like in the previous example, the string at the origin of the form of the metric {\em is not} the source of the KR field that we calculate.}
We can also write down the metric tensor 
\be
(g\dd\mu\nu)=\bpm
1&0&0&0\\
0&-1&0&0\\
0&0&-(\rho^2+\beta^2)&-\beta\\
0&0&-\beta&-1
\epm,
\ee
which allows  building the 
 dual tetrad
basis 
$\df e^0=cdt$, $\df e^1=d\rho$, $\df e^2=\rho d\varphi$ and $\df e^3=\beta d\varphi+dz$.

\Rvsd{Due to the use of the same system of local coordinates,} the 
1-form $\df T$  takes the same simple form as for the wiggly string,
$\df T=T_2(\rho) \df e^2$, and its line integral along the contour of radius $\rho$ is again transparent
\be
\int \df T=\int T_2 \ \!\df e^2=\int T_2(\rho)\rho d\varphi=2\pi\rho T_2(\rho)=\Phi_\sigma.
\ee
It follows the same expression (\ref{eq-79}) on the local coordinate basis.
Then we obtain
\bea
g_0\star \df T&=&g_0T_2\df e^3\wedge\df e^1\wedge\df e^0\nnb\\
&=&g_0\frac{\Phi_\sigma}{2\pi\rho}(dz\wedge d\rho-\beta d\rho\wedge d\varphi)\wedge cdt
\eea
and, as a result of the non-diagonal character of the metric,  the spacetime of the chiral cosmic string induces a non-zero vacuum polarization, with anisotropic,  space-dependent polarizability and two non-vanishing components for the 2-form $\df R$
\bea
&&R\dd z\rho=\frac{g_0}{\rho}{T_\varphi},\label{eqRzr}\\
&& R\dd \rho\varphi=-\frac{g_0\beta}{\rho}{T_\varphi}\label{eqRrf}.
\eea

\section{Discussion}
It is important to stress that symmetry (here cylindrical symmetry) dictates the form of  $\df T$, since the relation $d\df T=\upsigma$ is a topological relation (it does not involve the metric tensor), hence it does not depend on the precise choice of coordinates. $\df T$ has the same expression in the coordinate basis in the examples considered (see e.g. equations (\ref{eq-59}) and (\ref{eq-79})).
On the other hand, $\df R$ is obtained via  Hodge duality. This means that the precise form of the metric on the local basis plays a central role. This is not true in the Minkowskian frame (more precisely the metric tensor components, \Rvsd{there,} are constants) for which one always observes an absence of polarizability.

An interesting feature of the relations between $\df R$ and $\df T$ concerns the property of gauge invariance w.r.t the choice of the coframe. Indeed, the cotetrad basis vectors define a set of sixteen components, ${e^a}_\mu$ which obey the constraint
\be
{e^a}_\mu{e^b}_\nu\eta\dd ab=g\dd\mu\nu.
\ee
There are thus only ten relations among them because the metric tensor $g\dd\mu\nu$ is symmetric. This leaves a freedom, that is also called gauge freedom, in the orientation of the orthonormal basis (three rotations and three boosts which preserve the covariance of the laws of Special Relativity under Lorentz transformations). This property is ensured by 
the  properties of change of bases in a vector space, but 
it is instructive for our purpose to \Rvsd{study}   an example. Let us then consider the case of the  chiral cosmic string and, instead of the previous tetrad, we now propose another choice
$\tilde{\df e}^0=cdt$, $\tilde{\df e}^1=\cos\varphi d\rho-\rho\sin\varphi d\varphi$, 
$\tilde{\df e}^2=\sin\varphi d\rho+\rho\cos\varphi d\varphi$ and $\tilde{\df e}^3=(\beta/\rho)\sin\varphi d\rho+\rho\cos\varphi d\varphi+dz$. In terms of these, the 1-form $\df T$ now becomes $\df T=\tilde T_1\tilde{\df e}^1+\tilde T_2\tilde{\df e}^2$ with $\tilde T_1=-\sin\varphi\ \! T_2$ and $\tilde T_2=\cos\varphi \ \!T_2$ with no apparent symmetry. The basis $\{\df e^a\}$ is an orthonormalized ``cylindrical'' basis while $\{\tilde{\df e}^a\}$ is a ``Cartesian'' basis, both with Minkowski metrics. The former is adapted to the natural expression of cylindrical symmetry while the second is not.
Nevertheless, it is easy to show also that Eqs.~(\ref{eqRzr}) and (\ref{eqRrf}) remain true.

We have learned enough from the examples above to draw general conclusions
for arbitrary metrics (we will focus here on the use of local coordinates of ``cylindrical type'' $(ct,\rho,\varphi,z)$). We still consider cylindrically symmetric sources, and as we have argued earlier, this can have a sensible meaning only in the orthonormalized basis  $\df e^a={\df e^a}_\mu dx^\mu$ of the cotangent spacetime, so here $\df T=T_\mu dx^\mu=T_2(\rho)\df e^2$. Using the relation $\df e^2={e^2}_\mu dx^\mu$ between the orthonormalized and the local basis, and integrating in the former basis along a closed curve at $\rho$ and $z$ constant, one has 
\be
\int_{\cal C}\df T=T_2(\rho)\int_0^{2\pi} {e^2}_\varphi d\varphi=\Phi_\sigma\ee
or 
\be
T_2(\rho)=\Phi_\sigma \left(\int_0^{2\pi} {e^2}_\varphi d\varphi\right)^{-1},\quad
T_\varphi= {e^2}_\varphi T_2(\rho).
\ee
In the case where ${e^2}_\varphi$ does not depend on $\varphi$ (all examples studied earlier fall into this category, except the ``tilde'' rotated basis discussed above), this is simply $T_2(\rho)=\Phi_\sigma/(2\pi{e^2}_\varphi)$
and $T_\varphi=\Phi_\sigma/(2\pi)$.
Now, in the orthonormalized basis, we have calculated $\star \df T$ several times already and we pass to the local basis using $\df e^3={e^3}_\mu dx^\mu$ and $\df e^1={e^1}_\nu dx^\nu$. For the time component, there is a certain freedom and we set
$\df e^0={e^0}_tcdt+{e^0}_i dx^i$ to identify a particular time direction. This defines a general formula
\bea
 \df R&=&g_0\frac{T_\varphi}{{e^2}_\varphi}({e^3}_\mu{e^1}_\nu-{e^3}_\nu{e^1}_\mu){e^0}_tdx^\mu\wedge dx^\nu.
\eea
All previous results are recovered from this general expression.

One may also wonder whether we can easily generate a non-zero 3-form $\df K$. The discussion above shows that this will be the case if $\df e^0$ comprises space terms in ${e^0}_i dx^i$.
An example is the spinning cosmic string~\cite{perlick}
\be
ds^2=(cdt-ad\varphi)^2-d\rho^2-\alpha^2\rho^2d\varphi^2-dz^2
\ee
which couples the time coordinate to the angle of rotation (here around $z$) and
for which $\df T$ is still given by (\ref{eq-79}). The presence of $d\varphi$ in $\df e^0=cdt-ad\varphi$ then leads to two terms in $\star \df T$,
\bea
&&g_0\star\df T=\df R\wedge cdt+\df K\quad\hbox{with}\nnb\\
&&\qquad \df R=
g_0\frac{T_\varphi}{\rho}dz\wedge d\varphi\\
&&\qquad\df K=-g_0\frac{aT_\varphi}{\rho} d\rho\wedge d\varphi\wedge dz,
\eea
and, as announced, a non-zero $\df K$ emerges.

Other particular metrics not considered in this paper could also be of interest. Let us mention the case of Kleinian metrics with specific signature alterations leading to  the emergence of two time components~\cite{PhysRevA.92.063806,alves2021implications}. This may be of specific interest in the context of the identification of the 2-form $\df R$.

Finally, we wish to close this discussion on a point raised in the course of this article. The Kalb-Ramond field is a candidate to describe the torsion of spacetime. Here we have assumed various background spacetimes in which we have calculated the Kalb-Ramond fields, neglecting a possible back-reaction on the geometric structure of spacetime. An interesting extension of this work would be to consider for example the free motion of test particles in the original background spacetime modified by the induced torsion. In particular, due to the presence of torsion, the geodesic curves would now differ from the auto-parallel curves.

%\bibliography{references}

\end{document}